\documentclass[prd,twocolumn,showpacs,nofootinbib,preprintnumbers]{revtex4}

\usepackage{amsmath}
\usepackage{amsfonts}
\usepackage{graphicx}
\usepackage{dcolumn}
\usepackage{hyperref}

\def\be{\begin{equation}}
\def\ee{\end{equation}}
\def\ba{\begin{eqnarray}}
\def\ea{\end{eqnarray}}
\def\bs{\begin{subequations}}
\def\es{\end{subequations}}
\renewcommand{\S}{{\text{\tiny $\phi$}}}
\newcommand{\T}{{\text{\tiny $T$}}}

\begin{document}

\title{Consistency relations and degeneracies in (non)commutative patch inflation}
\author{Gianluca Calcagni}
\email{calcagni@fis.unipr.it}
\affiliation{Dipartimento di Fisica ``M. Melloni,'' Universit\`{a} di Parma, Parco Area delle Scienze 7/A, I-43100 Parma, Italy\\}
\affiliation{INFN -- Gruppo collegato di Parma, Parco Area delle Scienze 7/A, I-43100 Parma, Italy}

\date{June 4, 2004}

\begin{abstract}
The consistency equations of patch inflation are considered in a next-to-leading-order slow-roll (SR) expansion. Some general aspects of braneworld degeneracy are pointed out, both with an ordinary scalar field and a Born-Infeld tachyon. The discussion is then extended to the maximally symmetric noncommutative case.
\end{abstract}

\preprint{UPRF-2004-08}

\maketitle


\section{Introduction and setup}

In recent years new gravity models invoking high-energy physics and extra dimensions have been developed; in many of them, the visible 4D Universe is actually a 3-brane embedded in a five-dimensional bulk. Randall-Sundrum (RS) and Gauss-Bonnet (GB) scenarios are typical examples of this setup, where the projected equations on the brane give rise to an effective 4D evolution which takes at least part of the bulk physics into account. 

An important aspect of the inflationary early Universe is the emerging of a set of ``consistency'' relations involving some of the most relevant observables, that is the amplitudes and indices of the perturbation spectra generated by quantum fluctuations stretched outside the Hubble horizon during the accelerated expansion. These relations do not depend on the form of the inflationary potential but do depend on either the type of scalar field (ordinary or tachyonic) on the brane and the details of the high-energy geometrical model. The consistency equations are a typical result from inflation that other theories of structure formation are not able to reproduce and reflect the common physical origin of scalar and tensor perturbations; this scalar-tensor entanglement is even more pronounced in the braneworld framework.
 
In this Letter we address the issue of possible theoretical degeneracies between next-to-leading-SR-order consistency relations of different inflaton-braneworld models, to be distinguished from observational degeneracies coming from particular values of the observables, for example, when the spectrum is nearly scale invariant in the extreme SR approximation. This problem arose for the first time when the degeneracy of the 4D and RS lowest-SR-order tensor index was discovered \cite{HuL1,HuL2}; several works then showed that this degeneracy is most likely incidental (e.g. \cite{cal2} and references therein). We are going to confirm this result in quite a general manner and pave the way to the classification of eventual future braneworld scenarios. The nondegeneracy of the consistency relations is then substantially confirmed in the case of a noncommutative brane with a realization of the noncommutative algebra preserving the FRW symmetry; see \cite{BH,cal4} and references therein. The introduction of the new degree of freedom provided by a noncommutative parameter slightly complicates the analysis but does not trigger further degeneracies in any of the concrete (non)commutative braneworlds.

It is useful to stress that even in the standard general relativistic case the consistency relations are violated in some simple situations, for example, in multi-field inflation (see, e.g., \cite{WBMR} and references therein). In this sense, a deviation from the standard equations would not provide the smoking gun for the existence of extra dimensions.

According to the aim and degree of theoretical accuracy one wants to achieve, a number of technical approaches to high-energy models are at hand. As a first step toward a comprehension of the subject we choose the patch formulation of inflationary cosmology \cite{cal3}, in which the effective Friedmann equation from the point of view of a brane observer is
\be \label{FRW}
H^2=\beta_q^2 \rho^q\,,
\ee
where $q$ is constant and $\beta_q>0$ is some model-dependent proportionality factor; we do not need to specify $\beta_q$ since it will not appear in the slow-roll expressions for the cosmological observables. We neglect any contribution from both the Weyl tensor and the brane-bulk exchange; assuming there is some confinement mechanism for a perfect fluid with equation of state $p=w\rho$, the continuity equation on the brane reads $\dot{\rho}+3H (\rho+p) = 0$. We then consider an inflationary four-dimensional flat brane filled with an homogeneous inflaton field $\psi$ with potential $V$; this field can be either an ordinary scalar $\phi$, with energy density and pressure $\rho = \dot{\phi}^2/2 + V(\phi) = p+2V(\phi)$, or a ``tachyon'' field $T$ satisfying a Born-Infeld action and with energy density $\rho =V(T)/c_S$ and pressure $p= -V(T) c_S$, where $c_S=\sqrt{-w}=\sqrt{1-\dot{T}^2}$.

In order to write down the cosmological observables in terms of slow-roll parameters, we define the first parameter as the time variation of the Hubble length $H^{-1}$,
\be \label{epsilon}
\epsilon \equiv -\frac{\dot{H}}{H^2}=\frac{3q(1+w)}{2}\,.
\ee
For a scalar field, the first three SR parameters are
\bs\ba
\epsilon_\S &\equiv& \epsilon =\frac{3q\beta_q^{2-\theta}}{2} \frac{\dot{\phi}^2}{H^{2-\theta}}\,, \label{phepsilon}\\
\eta_\S     &=&  -\frac{\ddot{\phi}}{H\dot{\phi}} \label{eta}\,,\\
\xi^2_\S    &=&   \frac{1}{H^2} \left(\frac{\ddot{\phi}}{\dot{\phi}}\right)^.\,,\label{xi}
\ea\es
where $\theta \equiv 2(1-q^{-1})$. For the tachyon field,
\bs\ba
\epsilon_\T &\equiv& \epsilon = \frac{3q}{2} \dot{T}^2\,, \label{Tepsilon}\\
\eta_\T         &=&  -\frac{\ddot{T}}{H\dot{T}}\,, \\
\xi_\T^2  &=& \frac{1}{H^2} \left(\frac{\ddot{T}}{\dot{T}}\right)^. \,.\label{Txi}
\ea\es
The evolution equations of the parameters with respect to synchronous time are second-SR-order expressions,
\bs \label{dotSR}
\ba
\dot{\epsilon} &=& H\epsilon \left[\left(2-\widetilde{\theta}\right)\epsilon-2\eta\right]\,,\label{epsih'}\\
\dot{\eta}     &=&  H\left(\epsilon\eta-\xi^2\right)\,,\label{etah'}
\ea
\es
where $\widetilde{\theta}=\theta$ for the ordinary scalar and $\widetilde{\theta}=2$ for the tachyon. In expressions valid both for $\phi$ and $T$ we will specify no subscript for the SR parameters.

Table \ref{table1} shows the numerical values of the three main energy ``patches'' constituting the evolution of a GB braneworld, from high to low energy regimes. Here $\zeta_q(h)$ is a quantity determined by the specific gravitational-geometric configuration one is considering \cite{LMW,DLMS}, although from our point of view it will play the role of a purely numerical coefficient; it contributes to the normalization of the tensor amplitude through its general definition, Eq. (\ref{zgen}). By comparing Gauss-Bonnet and 4D scenarios, it is clear that the case $\zeta_q(h)=\zeta_{q'}(h)$ is possible even when $\theta \neq \theta'$. In the discussion on degeneracy we will not restrict ourselves to these three scenarios, in the perspective other gravity models might generate patches (i.e., effective Friedmann equations with $\theta$) different from $\theta = 0,\,\pm 1$.
\begin{table}[floatfix]
\caption{\label{table1} Gauss-Bonnet, Randall-Sundrum and 4D scenarios.}
\begin{ruledtabular}
\begin{tabular}{ccdc}
Regime &   $q$   &   \theta  &          $\zeta_q(h)$           \\ \hline
GB     &  $2/3$  &      -1   & 							1 							\\
RS     &    2    &       1   &      			$2/3$   				  \\
4D     &    1    &       0   &         			1        				\\
\end{tabular}\end{ruledtabular}
\end{table}


\section{Braneworld spectra}

By definition, the 4D spectral amplitude generated by the $k$th mode of the perturbation $\Phi$ is
\be \label{ampli}
A^2 \equiv \frac{2k^3}{25\pi^2} \left\langle |\Phi_k|^2\right\rangle\!\Big|_*\,,
\ee
where angle brackets denote vacuum expectation value and the expression is evaluated at the horizon crossing time defined by $k(t_*)=a(t_*)H(t_*)$. In the case of the scalar spectrum (subscript ``$s$''), $\Phi={\mathcal R}$ is the curvature perturbation on comoving hypersurfaces, generated by the scalar field filling the early Universe. For the gravitational spectrum (subscript ``$t$''), $\Phi$ denotes the coefficient functions of the zero mode  $h_{\mu\nu}^{(0)}(x)$ of the 4D polarization tensor.

Neglecting the contribution of the Weyl tensor and the total anisotropic stress, the system of equations closes on the brane and the number of gauge degrees of freedom conveniently reduces for longitudinal scalar perturbations; moreover, bulk effects are suppressed in the long wavelength limit, $k \ll aH$. In this case one can assume the validity of the 4D Mukhanov equation on the brane,
\be \label{muk}
u_k''+\left(k^2-\frac{z''}{z}\right)u_k=0\,,
\ee
where primes stand for second derivates with respect to conformal time 
\be \label{confor}
\tau\equiv\int \frac{dt}{a} = -\frac{1}{aH(1-\epsilon)}\,,
\ee
and $u_k$ are the coefficients of the plane wave expansion of the canonical variable $u=-z\Phi$. For a perfect fluid, the squared function $z$ is
\be \label{zgen}
z^2 =\zeta_q \frac{a^2 (\rho+p)}{H^2}= \zeta_q\frac{a^2 (1+w)}{\beta_q^{2-\theta}H^\theta}\,,
\ee
where $\zeta_q$ is a proportionality coefficient related to the field $\Phi$; for an ordinary and tachyon scalar on the brane, respectively,
\ba
z(\phi) &=& \frac{a\dot{\phi}}{H}\,,\\
z(T)    &=& \frac{a\dot{T}}{c_S\beta_q^{1/q} H^{\theta/2}}\,,
\ea
with $\zeta_q(\phi)=1$ and $\zeta_q(T)=1/c^2_S$. In the extreme SR approximation we can set $\zeta_q(T)\approx 1$. The amplitude (\ref{ampli}) becomes
\be
A^2 = \frac{2k^3}{25\pi^2}\frac{|u_k|^2}{z^2}\,;
\ee
note that the lowest-SR-order version of the scalar amplitude agrees with that obtained via a de Sitter calculation of the correlation function of the fluctuation $\delta\psi \propto u/a$ outside the horizon.

The gravitational spectrum in RS and GB scenarios has been investigated in \cite{LMW,DLMS} for a de Sitter brane, whose maximal symmetry permits a variable separation of the wave equation for the Kaluza-Klein gravity modes, $h_{\mu\nu}(x,y) \rightarrow h_{\mu\nu}^{(m)}(x)\xi_m(y)$, where $y$ is the extra direction.  The normalization of the bulk-dependent part of the graviton zero mode, calculated on the brane position $y_b$, determines the mapping function $F\equiv\xi_0(y_b) \kappa_5/\kappa_4$, where $\kappa_n$ is the gravitational coupling in $n$ dimensions. It turns out that $F$ is a complicated function of the couplings of the theory, the Hubble parameter $H$ and $\chi$, the inverse of the bulk curvature scale; given the 4D amplitude $A_{t(4D)} \propto \kappa_4 H$ with $z_{4D}=\sqrt{2}a/\kappa_4$, the braneworld tensor amplitude is $A_t=A_{t(4D)} F (H/\chi)$. To be consistent with the patch solution (\ref{FRW}),
we must consider the approximated version $F_q$ of $F$ in the proper energy limits.

Writing $A_t=A_{t(4D)} z_{4D}/z$, one may encode the phenomenology of the transverse direction into a map acting on the function $z$, $z_{4D} \mapsto z=z_{4D}/F_q$. We can find the patch version of $F$ with a trick, by noting that in four dimensions the graviton background can be formally described by Eq. (\ref{zgen}) with $\zeta_1(h)=1$ and a perfect fluid $p_h= -\rho_h/3$ which does not contribute to the cosmic acceleration, since $\ddot{a}=aH^2(1-\epsilon)$ and $\epsilon=1$. Generalizing this stationary solution one has $w_h=2/(3q)-1$ and \cite{cal4}
\ba
z(h)  &=& \frac{\sqrt{2}a}{\kappa_4 F_q}\,,\label{zgrav}\\
F^2_q &\equiv& \frac{3q\beta_q^{2-\theta}H^\theta}{\zeta_q(h)\kappa_4^2}\,.\label{Fgeneral}
\ea
This is equivalent to take the 4D tensor amplitude and substitute the gravitational coupling with $\kappa_4^2 \sim (H^2/\rho)_{4D} \rightarrow H^2/\rho$. Although these arguments do not completely justify Eq. (\ref{Fgeneral}) as the general patch solution for the tensor amplitude, the proposed ingredients provide a very compact notation that does match the results coming from both the 4D and full 5D calculations in Randall-Sundrum and Gauss-Bonnet scenarios. We suggest this picture to be valid in other cases, too; direct contact with explicit gravity models is reduced to a minimum through the coefficients $\beta_q$ and $\zeta_q$, but only the latter is indispensable for the consistency relations. The spectral indices and their runnings are defined as
\ba
n_t &\equiv& \frac{d \ln A_t^2}{d \ln k}\,,\qquad n_s-1 \equiv \frac{d \ln A_s^2}{d \ln k}\,,\\
\alpha &\equiv& \frac{d n}{d \ln k}\,.
\ea
If $z''/z \propto \tau^{-2}$, the Mukhanov equation is exactly solvable: setting
\be
\nu^2 \equiv \frac{1}{4}+\tau^2\frac{z''}{z}\,,
\ee
the large-scale solution, perturbed according to the SR approximation, is
\begin{eqnarray*}
|u_k| &=& \frac{2^{\nu-3/2}}{\sqrt{2k}}\frac{\Gamma(\nu)}{\Gamma(3/2)}\left(-k\tau\right)^{-\nu+1/2}\\
      &\approx& [1-C(\nu-3/2)]\frac{(-k\tau)^{-\nu+1/2}}{\sqrt{2k}} \,,\quad \nu-3/2 \ll 1\,,
\end{eqnarray*}
where $C=\gamma+\ln 2-2 \approx -0.73$ is a numerical constant ($\gamma$ is the Euler-Mascheroni constant).
Since $\nu$ is a combination of SR parameters, this expression describes a cosmological solution with constant SR parameters; patch power-law inflation has exactly this feature \cite{cal3} but there are other feasible solutions \cite{cal1,RL}.

Let us list the results for the cosmological spectra to next-to-leading SR order. Using Eq. (\ref{confor}) together with $\partial_\tau^2=a^2(H\partial_t +\partial_t^2)$, for the tensor amplitude we have
\ba
\frac{z''}{z} &=& (aH)^2\left[2+\left(\frac{3\theta}{2}-1\right)\epsilon\right]\,,\\
\nu_h &=& \frac{3}{2}+\left(1+\frac{\theta}{2}\right)\epsilon\,,
\ea
and
\bs \label{At}\ba
A_t^2 &=& \left\{1-\left[(2+\theta)C+2\right]\epsilon\right\}\frac{3q\beta_q^{2-\theta}}{25\pi^2}\frac{H^{2+\theta}}{2\zeta_q}\,,\label{Ah}\\
n_t &=& -(2+\theta)\epsilon\left[1+\omega_t\epsilon-2\left(C+\frac{2}{2+\theta}\right)\eta\right]\,,\nonumber\\ \\
\alpha_t &=& (2+\theta)\epsilon\left[2\eta-\left(2-\widetilde{\theta}\right)\epsilon\right],
\ea
where from now on $\zeta_q=\zeta_q(h)$ and
\be
\omega_t \equiv \left(2-\widetilde{\theta}\right)C+\frac{6-\widetilde{\theta}}{2+\widetilde{\theta}}\,.
\ee\es
The $O(\epsilon^2)$ part of the tensor index and its running depend on the assumed scalar field model through Eq. (\ref{epsih'}). In the tachyon case, $\omega_t= 1$. For the ordinary scalar field,
\ba
\frac{z''}{z} &=&  (aH)^2\left[2+2\epsilon_\S-3\eta_\S\right]\,,\\
\nu_\S  &=& \frac{3}{2}+2\epsilon_\S-\eta_\S\,,
\ea
and
\bs\label{AsS} \ba 
A_s^2(\phi) &=& [1-2(2C+1)\epsilon_\S+2C\eta_\S]\frac{3q\beta_q^{2-\theta}}{25\pi^2}\frac{H^{2+\theta}}{2\epsilon_\S}\,,\nonumber\\\label{phiS}\\
n_s-1 &=& \left(2\eta_\S-4\epsilon_\S\right)+2(5C+3)\epsilon_\S\eta_\S- 2C\xi_\S^2\nonumber\\
&&-2[(4-\theta)+ 2(2-\theta)C]\, \epsilon_\S^2\,,\\
\alpha_s &=& 2\left[2(\theta-2)\, \epsilon_\S^2+5\epsilon_\S\eta_\S-\xi_\S^2\right],\label{alp}\\
r &=&  \frac{\epsilon_\S}{\zeta_q}\left[1-(\theta-2)C\epsilon_\S-2C\eta_\S\right],
\ea\es
where $r \equiv A_t^2/A_s^2$. Finally, for the tachyon one gets
\ba
\frac{z''}{z} &=& (aH)^2\left[2+\left(\frac{3\theta}{2}-1\right)\epsilon_\T-3\eta_\T\right]\,,\label{tac1}\\
\nu_\T  &=& \frac{3}{2}+\left(1+\frac{\theta}{2}\right)\epsilon_\T-\eta_\T\,,\label{tac2}
\ea
and
\bs \label{AsT} \ba 
A_s^2(T) &=& \left(1-2\omega_s\epsilon_\T+2C\eta_\T\right)\frac{3q\beta_q^{2-\theta}}{25\pi^2}\frac{H^{2+\theta}}{2\epsilon_\T}\,,\label{tacS}\\
n_s-1 &=& \left[2\eta_\T-(2+\theta)\,\epsilon_\T\right]+2\left(C+1+2\omega_s\right)\epsilon_\T\eta_\T\nonumber\\ &&-2C\xi_\T^2-(2+\theta)\epsilon_\T^2\,,\\
\alpha_s &=& 2\left[(3+\theta)\,\epsilon_\T\eta_\T-\xi_\T^2\right],\label{Talp}\\
r &=& \frac{\epsilon_\T}{\zeta_q}\left[1-(2-\theta)\frac{\epsilon_\T}{6}-2C\eta_\T\right].
\ea
Here,
\be
\omega_s \equiv \left(C+\frac{5}{6}\right)+\frac{\theta}{2}\left(C+\frac{1}{6}\right)\,.\label{tac3} 
\ee\es

As a last remark, we note that there is a sort of triality among the Mukhanov equations for the scalar, tachyon and tensor amplitudes: in fact, $\nu_\S=\lim_{\theta\rightarrow 2}\nu_\T$ and $\nu_h=\lim_{\eta\rightarrow 0}\nu_\T$. The first condition is a consequence of the definitions of the SR towers \cite{cal3}, the second one states that, when $\epsilon_\T \propto \dot{T}^2 \approx \text{const}$, the quantum field $u_k(\delta T)$ evolves like its gravitational counterpart $u_k(h)$. One might find interesting to rewrite the SR expressions (\ref{AsS}) and (\ref{AsT}) for the scalar perturbation in a minimal fashion, by substituting $\theta$ with $\overline{\theta}=2+\theta-\widetilde{\theta}$ in Eqs. (\ref{tac1}), (\ref{tac2}) and (\ref{tac3}); in the ordinary scalar case, $\omega_s= 2C+1$. Then,
\bs\ba
A_s^2 &=& \left(1-2\omega_s\epsilon+2C\eta\right)\frac{3q\beta_q^{2-\theta}}{25\pi^2}\frac{H^{2+\theta}}{2\epsilon}\,,\\
n_s-1 &=& \left[2\eta-\left(2+\overline{\theta}\right)\,\epsilon\right]+2(C+1+2\omega_s)\epsilon\eta\nonumber\\ &&-\left[2\left(2-\widetilde{\theta}\right)\omega_s+2+\overline{\theta}\right]\epsilon^2-2C\xi^2\,,\\
\alpha_s &=& 2\left[2\left(\widetilde{\theta}-2\right)\epsilon^2 +\left(3+\overline{\theta}\right)\epsilon\eta-\xi^2\right]\,,\\
r &=& \frac{\epsilon}{\zeta_q}\left\{1-\left[(2+\theta)C+2(1-\omega_s)\right]\epsilon-2C\eta\right\}.\nonumber\\
\ea\es
If $\xi \ll \min(\epsilon,|\eta|)$, we can collect Eqs. (\ref{At}), (\ref{AsS}) and (\ref{AsT}) in the closed set of consistency equations:
\bs\ba
\alpha_s(\phi) &\approx& \zeta_q r [4(3+\theta)\zeta_qr+5(n_s-1)]\,, \label{cealpsS}\\
\alpha_s(T) &\approx& (3+\theta)\zeta_q r [(2+\theta)\zeta_qr+(n_s-1)]\,, \label{cealpsT}\\
n_t(\phi) &=& \zeta_q r [-(2+\theta)+(2+\theta)\zeta_qr+2(n_s-1)]\,,\nonumber\\\label{cenS}\\
n_t(T) &=& \zeta_q r \left[-\vphantom{\frac{1}{1}}(2+\theta)+2(n_s-1)\right.\nonumber\\
&&\left.\qquad\quad+\frac{(2+\theta)(4+\theta)}{6}\zeta_qr\right],\label{cenT}\\
\alpha_t &=& (2+\theta)\zeta_qr[(2+\theta)\zeta_qr+(n_s-1)]\,.\label{cealpt}
\ea\es
The consistency equation for the tensor index depends on the chosen scalar action, as first pointed out in \cite{SV}. Conversely, Eq. (\ref{cealpt}) is valid both for the scalar and tachyon field, but in general it is model-dependent \cite{SV}. The next-to-lowest-order equations (\ref{cenS}) and (\ref{cenT}) generalize the lowest-order equation presented in \cite{cal4}.


\section{Degeneracy of consistency equations}

Let us now seek what are the necessary conditions for obtaining the same set of consistency equations in two models $(\psi,\,\theta)$ and $(\psi',\,\theta')$. There are several possible degeneracies which arise particular attention. The first one is \emph{exact}, that is $\alpha_s=\alpha_s'$, $\alpha_t=\alpha_t'$ and $n_t=n_t'$ to next-to-leading SR order; this model correspondence would open up many compelling possibilities, for example to construct a complicated braneworld scenario starting from a simple one. A second, more operative degeneracy is \emph{effective}, namely, one considers only the scalar running and the lowest-SR-order tensor index. Differences in next-to-leading-order tensor indices and in tensor runnings are neglected since the observational uncertainty on these quantities would blur any theoretical mutual deviation, at least for near-future experiments. When neither exact nor effective degeneracy are achieved, we will say that the two classes of models are \emph{definitely} nondegenerate. Another choice could be to consider \emph{tensor} degeneracy, of either lowest and next-to-lowest order, when the tensor index and its running are degenerate; tensor-degenerate models give the same gravitational wave spectrum. This degeneracy is useful when reducing the space of parameters in numerical analyses via the tensor-index consistency equation. 

The first degeneracy we investigate is between ordinary-scalar and tachyon-field scenarios. Let a prime denote the tachyon model; in order to get $(\phi,\,\theta)=(T,\,\theta')$, we match Eqs. (\ref{cealpsS}) and (\ref{cealpsT}), giving
\bs \label{deg1}
\ba
\theta &=& \frac{14+13\theta'}{4(3+\theta')}\,,\qquad \theta'\neq -3\,,\\
\zeta_q &=& \frac{3+\theta'}{5}\zeta_{q'}\,.
\ea
\es
From Eqs. (\ref{cenS}) and (\ref{cenT}) one gets either $\theta=2=\theta'$ or $\theta=-2=\theta'$; Eq. (\ref{cealpt}) is automatically degenerate for all $\theta$. Therefore, exact degeneracy is not allowed for finite $q$. For the effective degeneracy it is sufficient that 
\be \label{lSRce}
(2+\theta)\zeta_q=(2+\theta')\zeta_{q'}\,,
\ee
from the lowest-order tensor indices; coupling this condition with Eq. (\ref{deg1}) again gives $\theta=2=\theta'$. Therefore, $\phi$- and $T$-models are definitely nondegenerate for finite $q$.

Scalar models in different braneworlds are definitely nondegenerate, $(\phi,\,\theta)\neq(\phi,\,\theta')$, since it must be $\theta=\theta'$ in the scalar running. The same conclusion holds for tachyon models, $(T,\,\theta)\neq(T,\,\theta')$ if $\theta\neq\theta'$.

Tensor degeneracy is straightforward: all the previous models are tensor degenerate to lowest SR order when Eq. (\ref{lSRce}) holds. In particular, (1) scalar and tachyon scenarios in a given patch and (2) 4D and RS models  are tensor-degenerate at lowest order. Models with the same inflaton field $\psi$ and $\zeta_q=\zeta_{q'}$ are not tensor-degenerate; obviously, 4D and GB scenarios are not tensor-degenerate. Next-to-leading-order tensor degeneracy is possible only between $(\phi,\,-2)$ and $(T,\,-2)$, when
\bs\ba
\alpha_s(\phi) &\approx& \zeta_{1/2} r [4\zeta_{1/2}r+5(n_s-1)]\,, \label{cealpsS-2}\\
\alpha_s(T) &\approx& \zeta_{1/2} r (n_s-1)\,, \label{cealpsT-2}\\
n_t &=& 2\zeta_{1/2} r(n_s-1)\,,\label{cen-2}\\
\alpha_t &=&0\,.\label{cealpt-2}
\ea\es
As far as the author knows, no braneworld gravity model giving an effective Friedmann equation with $q=1/2$ has been developed so far. Table \ref{table2} summarizes the various degeneracies for finite $q$.
\begin{table}[ht]
\caption{\label{table2} Commutative patch degeneracies for finite $q$. ``n.t.l.'' stands for next-to-lowest order.}
\begin{ruledtabular}
\begin{tabular}{l|cc}
Degeneracy               & $(\phi,\,\theta)-(T,\,\theta')$ & $(\psi,\,\theta)-(\psi,\,\theta')$ \\\hline
Exact                    &                $-$              &                $-$                 \\
Effective                &                $-$              &                $-$                 \\
Tensor n.t.l. SR &         $\theta=\theta'=-2$     &                $-$                 \\
Tensor lowest SR         & Eq. (\ref{lSRce})  &  Eq. (\ref{lSRce}) with $\theta \neq \theta'$\\ 
\end{tabular}\end{ruledtabular}
\end{table}

We can extend the discussion to noncommutative models. The introduction of the fundamental string energy scale $M_\mathrm{s}$ generalizes the lowest-order consistency equations \cite{cal4}:
\bs\ba \label{noncom}
\alpha_s(\phi) &\approx& r \zeta_q \{(5-\sigma)(n_s-1)\nonumber\\
&&+[4(3+\theta)-\sigma (7+\theta-\sigma)]r \zeta_q\},\label{salph}\\
\alpha_s(T) &\approx& r \zeta_q \{(3+\theta-\sigma)(n_s-1)\nonumber\\
&&+[(2+\theta)(3+\theta)-\sigma (5+2\theta-\sigma)]r \zeta_q\},\label{talph}\\
n_t &=& [\sigma-(2+\theta)+O(\epsilon^2)] \zeta_q r\,,\\
\alpha_\mathrm{t} &\approx& (2+\theta-\sigma)r \zeta_q \left[(n_s-1)+
(2+\theta-\sigma)r \zeta_q\right].\nonumber\\\label{tenalph}\ea\es
Here, $\sigma$ is a nonnegative function encoding the dependence on the string scale; in the commutative limit or UV region of the spectrum, it is $\sigma \propto (H/M_s)^4$, where $H$ is evaluated at horizon crossing. In the far IR regime, one finds that either $\sigma=6$ or $\sigma=2$ for class 1 (FRW-sphere factored out) and class 2 (one single effective scale factor) models, respectively. Since theoretical degeneracy of consistency relations should be independent of the particular value of horizon-crossing quantities, we investigate only the IR case with constant $\sigma$; indeed, Eqs. (\ref{salph})--(\ref{tenalph}) are valid just in this approximation. We do not care for exact degeneracy and consider only effective and tensor degeneracy between two models $(\psi,\,\theta,\,\sigma)$ and $(\psi',\,\theta',\,\sigma')$, always discarding the commutative case $\sigma =\sigma'=0$. Patch models display lowest-order tensor degeneracy when 
\be \label{SRcenc}
(2+\theta-\sigma)\zeta_q=(2+\theta'-\sigma')\zeta_{q'}\,.
\ee
If $\sigma\neq\sigma'$, this translates into $\sigma_{4D}-\sigma_{GB}=1$, $\sigma_{4D}=2\sigma_{RS}/3$ or $2\sigma_{RS}/3-\sigma_{GB}=1$. As regards effective degeneracy:
\begin{itemize}
\item $\phi \leftrightarrow T$: Scalar and tachyon models are never degenerate;
\item $\phi \leftrightarrow \phi$: Degeneracy is possible only when $\sigma \neq \sigma'$ and at least one of the two models is not 4D, RS or GB. In general, it must be $\theta=(3\sigma'+10\theta'-2\sigma'\theta'-15)/(2-\theta')$, $\sigma=(\sigma'-5\theta'+5)/(2-\theta')$ and $\zeta_q=\zeta_{q'}(2-\theta')$;
\item $T \leftrightarrow T$: The degeneracy conditions read $\theta-\sigma=\theta'-\sigma'$ and $\zeta_q=\zeta_{q'}$. Therefore, $\sigma_{4D}-\sigma_{GB}=1$.
\end{itemize}
To conclude, among the known braneworld commutative models there are two lowest-SR-order tensor degeneracies, one between scalar and tachyon cosmologies and one between the four-dimensional scenario and the Randall-Sundrum braneworld; when noncommutativity is turned on, these braneworld models can be degenerate with suitable values of the noncommutative parameter, but not in the classes investigated in \cite{cal4}. This result holds under the standard assumption $\xi \ll \min(\epsilon,|\eta|)$, which permits to close the expression for the scalar running. If the inflaton potential does not satisfy such a dynamical constraint, as in the case of power-law ordinary inflation, then the consistency relations (\ref{salph}) and (\ref{talph}) are modified. For example, it may be convenient to perform numerical analyses via the horizon flow formalism; when one neglects the third flow parameter $\epsilon_3$ with respect to $\epsilon_1=\epsilon$, it turns out that the scalar and tachyon scenarios are always effectively degenerate when Eq. (\ref{SRcenc}) holds, since the scalar running is then given by Eq. (\ref{tenalph}) in both cases \cite{CT}. The discussion for the lowest-order tensor degeneracy would thus also apply to the effective degeneracy. Should this be the most realistic scenario, the adoption of one field instead of the other would be important only at second order in this model-independent context, in which nothing about the shape of the potential is said; anyway, the universal equation (\ref{SRcenc}) would continue to determine the condition for degeneracy, excluding coincident predictions from the braneworld (non)commutative setups we have considered.


\end{document}